\pdfoutput=1
\documentclass[12pt]{article}

\usepackage{amsmath,graphicx}
\usepackage{amsfonts}
\usepackage{amssymb}
\usepackage[nosort]{cite}
\usepackage{setspace}

\def\bk{{\bf k}}

\def\CO{{\cal O}}

\def\mpl{M_{\rm P}}
\def\half{\frac{1}{2}}

\makeatletter
\renewcommand\section{\@startsection {section}{1}{\z@}%
                                 {-3.5ex \@plus -1ex \@minus -.2ex}%
                                   {2.3ex \@plus.2ex}%
                                   {\normalfont\large\bfseries}}
\renewcommand\subsection{\@startsection{subsection}{2}{\z@}%
                                   {-3.25ex\@plus -1ex \@minus -.2ex}%
                                     {1.5ex \@plus .2ex}%
                                     {\normalfont\bfseries}}
\renewcommand\subsubsection{\@startsection{subsubsection}{3}{\z@}%
                                   {-3.25ex\@plus -1ex \@minus -.2ex}%
                                     {1.5ex \@plus .2ex}%
                                     {\normalfont\itshape}}
\makeatother

\newcommand{\Letter}{
\setlength{\textwidth}{16.5cm}
   \setlength{\textheight}{22.6cm}
    \hoffset=-0.5in
\voffset=-2.1cm }

\Letter

\newcommand{\be}{\begin{equation}}
\newcommand{\ee}{\end{equation}}
\newcommand{\bea}{\begin{eqnarray}}
\newcommand{\eea}{\end{eqnarray}}
\newcommand{\barr}{\begin{array}}
\newcommand{\earr}{\end{array}}
\def\bal#1\eal{\begin{align}#1\end{align}}

\begin{document}
\thispagestyle{empty}
\begin{flushright}
\end{flushright}
\vspace*{0.3in}
\begin{spacing}{1.1}

\begin{center}
{\large \bf Non-Bunch-Davies Anisotropy}

\vspace*{0.5in} {Xingang Chen$^{1}$ and Yi Wang$^{2}$}
\\[.3in]
{\em
$^1$Centre for Theoretical Cosmology, DAMTP,
\\University of Cambridge, Cambridge CB3 0WA, UK\\
$^2$Kvali Institute for the Physics and Mathematics of the Universe (WPI),
\\Todai Institutes for Advanced Study, University of Tokyo,
\\5-1-5 Kashiwanoha, Kashiwa, Chiba 277-8583, Japan
} \\[0.3in]
\end{center}

\begin{center}
{\bf
Abstract}
\end{center}
\noindent
We introduce a generic mechanism that can extend the effects of relic anisotropies at the beginning of inflation to relatively much shorter scales in density perturbations. This is induced by non-Bunch-Davies states of the quantum fluctuations, and can show up in the non-oscillatory components of the density perturbations. This mechanism works for general forms of anisotropies, and, to illustrate it, we use an example of relic vector field. The detailed scale-dependence of these anisotropies can be used to probe the initial quantum state of our universe.

\vfill
\newpage
\setcounter{page}{1}
\newpage

\section{Introduction}
\setcounter{equation}{0}

Generically, relic anisotropies at the onset of inflation will be swept away quickly. This is one of the successes of the inflation paradigm \cite{Guth:1980zm}. If inflation only lasted for the minimum amount of e-folds to solve the horizon and homogeneity problems, we may be able to see signatures of these initial relics on the largest scales in the sky. Such signatures should all have strong scale dependence,
but the details of this scale dependence are of crucial importance.

A naive expectation would be that, the anisotropies in the density perturbations, towards smaller scales, should decay as fast as the background relic anisotropies swept away by the inflation. For example, the background density of a relic massless vector field \cite{Chen:2013eaa} decays as fast as $1/a^4$, where $a$ is the scale factor of the universe. One would expect that the scale-dependence of the anisotropy in the density perturbations decay as fast as $1/k^4$, where $k$ is the comoving momentum of the density perturbation.

However, there is a caveat in this naive expectation. The nature of a final state is determined by not only the dynamics but also the initial condition. They play important roles in different components of the density perturbations.
\begin{itemize}
\item \textit{Dynamics}: The dynamics of the density perturbations, especially the frozen behavior around the horizon crossing, is controlled by the inflationary background in which the perturbations live. Thus, some anisotropic components in the perturbations, whose time-dependence is induced by the background evolution, should decay very fast in a speed dictated by the background dilution speed of the relics. This induces a corresponding scale-dependence for the anisotropy in density perturbations.  This is the underlying reason for the above expectation, and is verified in the example of \cite{Chen:2013eaa} if the Bunch-Davies (BD) state is chosen as the initial state of quantum fluctuations.
\item \textit{Initial condition}: Before inflation has settled to its attractor, the initial condition for the quantum fluctuations can be very model-dependent; and the BD state is not a natural choice.\footnote{The BD state is not the ground state for modes with finite wavelength at the beginning of inflation; and in addition, the initial state do not even have to be in the ground state. So, here the non-BD quantum states arise naturally not due to the trans-Planckian physics \cite{Brandenberger:2012aj}, but due to the initial conditions because we have a short period of inflation. Another difference from the trans-Planckian physics \cite{Brandenberger:2012aj} is that we will concentrate on the non-oscillatory components of the density perturbations, for reasons that we will explain later.}
    Phenomenologically the only restriction is that the state should not have infinite energy density in the small wavelength limit.
    So, in sufficiently short scales, the quantum state should approach the ground state of the Minkowski spacetime, which coincides with the BD vacuum.
    However, the scale-dependence that quantifies the ``sufficiently short scales'' in the above statement is independent of the background dilution speed.
    This scale-dependence is crucial in determining the scale-dependence in the time-independent coefficients in quantum fluctuations, which contribute to the final density perturbations; not just the background dilution speed. So the BD state is only a special case.

\end{itemize}

Now we can understand why the scale-dependence of the anisotropies should be sensitive to how we set the initial quantum state.
At the beginning of the inflation $\tau_0$ when the background is still anisotropic to a certain degree, the quantum fluctuations of fields in this background also inherit this degree of anisotropy. This applies to quantum fluctuations with {\em all} wavelengths, not just those near the horizon.
Therefore, the anisotropy should be imprinted in the coefficients of the quantum state with a scale-dependence that is more sensitive to the physics at $\tau_0$ than the late-time dilution. This is the source of the non-BD anisotropy in the density perturbations.
For realistic and well-motivated initial states, both the dynamics and the initial condition will play roles in determining the final density perturbations; and this is enough to give rise to qualitatively different scale dependence.
We emphasize that these anisotropies in the final density perturbations still have to decay away in sufficiently short scales, because as explained very short modes have to start their lives in the ground state; but their scale-dependence can be much weaker than that dictated by the background dilution speed, and even carry interesting features.
In the rest of this paper, we explain this mechanism through a detailed example.

This mechanism opens up a wider observational window on the initial conditions of inflation, and can have applications in a variety of anomalies related to anisotropies observed in WMAP and Planck at large scales \cite{Bennett:2010jb}, such as the planar asymmetry \cite{Tegmark:2003ve} and the hemispherical asymmetry \cite{Eriksen:2003db}.

\section{Relic vector field: UV-safe expansion}
\setcounter{equation}{0}

In this paper, we will use a specific model to illustrate this mechanism, namely the model of relic vector field \cite{Chen:2013eaa}.
In this example, we consider the general effects of the vector field on the inflationary background and the density perturbations. To study such effects, in the action
\bea
S=\int d^4x \sqrt{-g}
\left[ \frac{\mpl^2}{2} R - \frac{1}{4} F_{\mu\nu} F^{\mu\nu}
- \half \partial_\mu \phi \partial^\mu \phi - V(\phi) \right] ~,
\eea
we turn off any direct coupling between the vector field and the inflaton field $\phi$. The background vector field is chosen to be
\bea
F_{03}=-F_{30}=E(t) ~.
\eea
At the leading order in $E(t)$, the vector field dilutes with time as
\begin{align}
  E(t) = E_0 / a ~,
\end{align}
where $a\equiv \exp(H_0 t)$ is the scale factor of the attractor de Sitter background. Here $H_0 \equiv \sqrt{V_0/3}$ and $E_0$ are constants.

The vector field induces anisotropy in the metric. This in turn leads to a scale-dependent anisotropic component to the density perturbation, which is determined by the following equation of motion for the mode function $u_\bk$ of the inflaton fluctuation,  with the comoving momentum $\bk$,
\begin{align} \label{eq:eom-pert}
  \frac{1}{a^2} \frac{d}{d\tau}
  \left[ a^2 ( 1-\frac{E_0^2}{16 a^4H_0^2} ) u_\bk' \right]
  + (1-\frac{3 E_0^2}{16 a^4H_0^2}) (k_x^2 + k_y^2) u_\bk
  + (1+\frac{5 E_0^2}{16 a^4H_0^2}) k_z^2 u_\bk = 0 ~,
\end{align}
where $\tau$ is the conformal time defined as $\tau\approx -1/a H_0$, and the prime denotes derivative to $\tau$.
We set $\mpl=1$.
Solving this equation perturbatively in $\CO(E_0^2)$, the mode function $u_\bk$ with only the positive frequency mode (defined as the BD case) is given by
\bea
u_\bk = C_+ u_{\bk(0)}
+ \frac{H_0^3}{\sqrt{2k^3}} E_0^2 \sum_{n=3}^6 \alpha_{n} \tau^n e^{-ik\tau}
~,
\label{u_expand}
\eea
where
\bea
u_{\bk(0)} &=& \frac{H_0}{\sqrt{2k^3}} (1+ik\tau) e^{-ik\tau} ~,
\\
\alpha_{3} &=& -\frac{i}{24 k} (3-7 \cos^2\theta) ~,
\nonumber
\\
\alpha_{4} &=& \frac{1}{24} (3-7 \cos^2\theta) ~,
\nonumber
\\
\alpha_{5} &=& \frac{i k}{40} (3-7 \cos^2\theta) ~,
\nonumber
\\
\alpha_{6} &=& - \frac{k^2}{80} (1-4 \cos^2\theta) ~,
\eea
where $\theta$ is the polar angle from the direction of the vector field ${\bf E}$. The coefficient
\bal
|C_+|^2
= 1- \frac{1}{8}(3-7 \cos^2\theta) \frac{E_0^2 H_0^2}{k^4}
\label{C+}
\eal
determines the anisotropic component of the power spectrum. As mentioned in the Introduction, this anisotropic component decays very quickly towards the shorter scales, $\sim 1/k^4$, in accordance with the decay speed $1/a^4$ of the density of the massless vector field in the inflationary background.

Before generalizing this result to the non-BD case and properly discussing the physical picture we mentioned in the Introduction, let us first improve the expansion scheme of this solution. In this paper we would like to consider both the near-horizon modes and the modes well inside the horizon, so we examine the UV behavior of this solution more closely.
At large $k$, the leading order isotropic term in \eqref{u_expand} is proportional to $k\tau e^{-ik\tau}$, while the second order anisotropic term is $(H_0^2 E_0^2 \tau^4) k^2\tau^2 e^{-ik\tau} $.
So the degree of anisotropy seems to grow with $k$ and the perturbative expansion breaks down for large $k$-modes with $|k\tau| \gtrsim H_0^{-2}E_0^{-2}\tau^{-4}$. However, physically we still expect a well defined expansion scheme to exist for large momenta, because the effect of the vector field is parameterized by $E_0^2 H_0^2 \tau^4$, and when this parameter is small, all modes should receive a fraction of anisotropy of the same order. To improve this situation, we consider an alternative split between the zeroth and first order solution, and redefine the variable as
\begin{align}
  \psi_\mathbf{k}(\tau) \equiv \log u_\mathbf{k}(\tau)~.
\end{align}
We expand $\psi_\mathbf{k}$ in orders of $E_0^2$,
\begin{align} \label{eq:psi-pert-def}
  \psi_\mathbf{k} = \psi_{\mathbf{k}(0)} + \psi_{\mathbf{k}(1)} + \cdots ~.
\end{align}
The perturbative equation of motion takes the form
\begin{align} \label{eq:eom-pert-a2}
  \psi_{\bk(1)}'' + \psi_{\bk(1)}'^2 - 2\frac{\psi_{\bk(1)}'}{\tau} +k^2 - \frac{E_0^2H_0^2\tau^4}{8}
  \left[
    2 \frac{\psi_{\bk(0)}'}{\tau} + k^2 - 4 k_z^2  \right] =0 ~,
\end{align}
where the effect of the expansion parameter is more clear.

For comparison, we first solve for the BD case, where
\begin{align}
  \psi_{\mathbf{k}(0)} = \log u_{\mathbf{k}(0)} ~.
\end{align}
Inserting $\psi_{\mathbf{k}(0)}$ into \eqref{eq:eom-pert-a2}, we can solve $\psi_{\mathbf{k}(1)}$. Using the canonical quantization condition to fix the integration constant, we get
\begin{align} \label{eq:psi-expand}
  \psi_{\mathbf{k}(1)} = \frac{H_0^2 E_0^2}{(1+ik\tau)} \sum_{n=0}^6 \beta_n \tau^n~,
\end{align}
where
\begin{align}
  \beta_0 & = - \frac{1}{16 k^4} \left( 3-7\cos^2\theta \right)~,
  \nonumber \\
  \beta_1 & = -\frac{i}{16 k^3} \left( 3-7\cos^2\theta \right)~,\\
  \beta_2 & = 0~, \qquad \beta_m = \alpha_{m} ~(m=3,4,5,6)~.
\end{align}
Note that $\beta_0$ and $\beta_1$ reassemble the perturbative part of $C_+$, thus we have essentially re-summed the perturbative part of the solution \eqref{u_expand} into an exponential.

To generalize the solution to the non-BD case, we simply take the linear combination of two independent solutions,
\begin{align}
  u_\mathbf{k} =  C_{+0} e^{\psi_\mathbf{k}} + C_{-0} e^{\psi^*_\mathbf{k}}
  = C_{+0} u_{\bk(0)} e^{\psi_{\mathbf{k}(1)}} + C_{-0} u^*_{\bk(0)} e^{\psi^*_{\mathbf{k}(1)}} ~.
\end{align}
The normalization of $C_{+0}$ and $C_{-0}$ is again determined by the canonical quantization condition as in \cite{Chen:2013eaa},
\begin{align}
  \left| C_{+0} \right|^2 - \left| C_{-0} \right|^2 = 1~.
\label{C0_normalization}
\end{align}
Note that the definition of $C_{+0}$ here and $C_+$ in \eqref{u_expand} are different, even for the BD case.

Comparing to \eqref{u_expand}, this expansion scheme is much healthier in its UV behavior. For large $k$, the leading order $\psi_{(0)}$ is proportional to $ik\tau$, and the second order anisotropic term is $(E_0^2 H_0^2 \tau^4) k\tau$. Therefore, as we discussed in the Introduction, each mode, no matter how large its momentum is, receives an anisotropic correction of the same order. For the BD state, $\left| C_{+0}\right|=1$; after taking the late time limit $\tau\to 0$, most of these anisotropic terms vanish, leaving only the first two time-independent terms with the scale dependence $\sim 1/k^4$. In this case, the anisotropies originally present in the much higher momentum modes do not get a chance to show up in the density perturbations.

However if we consider the non-BD state, the normalization condition \eqref{C0_normalization} becomes non-trivial. It needs to be imposed at some early epoch $\tau_0$, during which the anisotropic components for all modes are still present. Generically, the same orders of anisotropies will be imprinted in the coefficients $C_{+0}$ and $C_{-0}$ at $\tau_0$, with a $k$-dependence arbitrary in principle and characteristic in well-motivated examples, as long as the backreaction is under control. These coefficients are time-independent. Non-BD vacua memorize primordial anisotropies in this fashion.

\section{An example: Gaussian state}
\setcounter{equation}{0}

In this section, we use a specific example of non-BD state to illustrate our mechanism. In \cite{Chen:2013eaa}, two numerical examples of non-BD states, the instant positive frequency state and the lowest energy state, are computed. There are two types of scale-dependence: one is oscillatory as $\sim \sin(k\tau_0)$, and another is non-oscillatory. In the numerical examples, the latter is swamped by the former and does not show up very clearly. However we expect their roles in the real observations to be opposite. Because the scales we are observing today are all near or within the horizon at $\tau_0$, this type of oscillatory running has a very large frequency because $\tau_0$ is around the beginning of the inflation.
After projecting to the multipole space, such signals are close to or beyond the ultimate resolution of the CMB experiments.

On the other hand, the non-oscillatory component is easier to observe. To clearly see the effect of this mechanism on the non-oscillatory component, it is better to use an analytical example. We choose the example of the Gaussian state \cite{Polarski:1995jg}, which is well motivated, easily tractable analytically and suffices to illustrate the main idea. Nonetheless, it is not necessarily the most natural initial state; other choices may well be possible. In fact, the experimental application of this mechanism can be the opposite: we can probe the initial state using properties of anisotropies from observations.

To start we write the quadratic Hamiltonian for the quantum fluctuation of the inflaton field, $\delta\phi$, in a canonical form.
Using the variable
\bea
v_\bk = z \delta\phi_\bk ~, \quad
z\equiv a \left( 1- \frac{E_0^2}{16a^4 H_0^2} \right)^{1/2} ~,
\eea
and its momentum conjugate
\bea
\pi_\bk = v'_\bk - \frac{z'}{z} v_\bk ,
\eea
the Hamiltonian is
\bal
H_2 = \int \frac{d^3 \bk}{(2\pi)^3}
\half \left[
\pi_\bk \pi_\bk^*
+ \hat k^2 v_\bk v_\bk^*
+\frac{z'}{z} ( v_\bk \pi_\bk^* + v_\bk^* \pi_\bk )
\right] ~,
\eal
where
\bal
\hat k^2 \equiv (1- \frac{E_0^2}{8a^4 H_0^2}) (k_x^2+k_y^2)+ (1+ \frac{3E_0^2}{8a^4 H_0^2}) k_z^2 ~.
\eal
The fields can be quantized either in the Heisenberg picture,
\begin{align}
v_\bk =& \frac{1}{\sqrt{2 \hat k}}
\left( a_\bk(\tau) + a_{-\bk}^\dagger(\tau) \right) ~,
\nonumber
\\
\pi_\bk =& -i \sqrt{\frac{\hat k}{2}}
\left( a_\bk(\tau) - a_{-\bk}^\dagger(\tau) \right) ~,
\label{Quan_Heisenberg}
\end{align}
or in the Schrodinger picture,
\begin{align}
v_\bk =&  f_\bk(\tau) a_\bk(\tau_0) + f_\bk^*(\tau) a_{-\bk}^\dagger(\tau_0) ~,
\nonumber
\\
\pi_\bk =& -i \left[ g_\bk(\tau) a_\bk(\tau_0) - g_\bk^*(\tau) a_{-\bk}^\dagger(\tau_0) \right] ~,
\label{Quan_Schrodinger}
\end{align}
where
\bea
f_\bk(\tau) = C_{+0} z e^{\psi_\bk} + C_{-0} z e^{\psi^*_\bk}
\eea
is proportional to the mode function and
\bal
g_\bk(\tau) = i (f'_\bk - \frac{z'}{z} f_\bk) ~.
\eal
The creation and annihilation operators in these two pictures satisfy the usual commutation relations, and are related to each other by the Bogolubov transformation.

The Gaussian state at $\tau_0$ is defined as
\bea
a_\bk(\tau_0)|0,\tau_0\rangle=0 ~.
\label{Def_G_state}
\eea
From \eqref{Quan_Heisenberg}, this means
\bea
(\pi_\bk - i \hat k v_\bk)|_{\tau_0} =0 ~.
\label{Gaussian_cond1}
\eea
From \eqref{Quan_Schrodinger}, we get
\bea
(g_\bk - \hat k f_\bk)|_{\tau_0} = 0 ~.
\label{Gaussian_cond2}
\eea
From \eqref{Gaussian_cond1}, we can see that this state has a Gaussian wave-functional, with a minimum uncertainty at $\tau_0$ \cite{Polarski:1995jg}.
Note that, in the case of the attractor isotropic inflation, the Gaussian state (\ref{Def_G_state}) becomes the BD state in the $\tau_0\to -\infty$ limit. However there is no such a simple relation between these two states for the anisotropic case. If we fix $E_0$, the limit $\tau_0 \to -\infty$ does not exist, since we are already at the beginning of the inflation with a finite $\tau_0$. We may instead fix $E_0^2 H_0^2 \tau_0^4$ and take the $\tau_0\to -\infty$ limit, by which we are looking at the deep subhorizon modes at $\tau_0$. This limit is the trivial isotropic inflation limit with BD vacuum.

The condition \eqref{Gaussian_cond2} gives a relation between $C_{+0}$ and $C_{-0}$; together with the normalization condition (\ref{C0_normalization}), they record anisotropy in these coefficients.
The power spectrum is proportional to the following quantity:
\bal
\left| C_{+0} + C_{-0} \right|^2 e^{2 \psi_{\bk (1)}|_{\tau\to 0}}
=& 1 - \frac{E_0^2 H_0^2}{8 k^4} (3-7\cos^2\theta)
\nonumber \\
& + \frac{1}{2k^2 \tau_0^2} +
E_0^2H_0^2\tau_0^4
\left( \frac{5}{16k^2\tau_0^2} - \frac{3}{8k^4\tau_0^4} - \frac{3}{16k^6\tau_0^6} \right)
\nonumber \\
& + E_0^2H_0^2\tau_0^4
\left( -\frac{3}{4k^2\tau_0^2} + \frac{7}{8k^4\tau_0^4} + \frac{7}{16 k^6\tau_0^6} \right) \cos^2\theta
\nonumber \\
& + {\rm oscillation ~ terms} ~,
\label{Csum_squared}
\eal
where the oscillation terms are proportional to $\sin(2k\tau_0)$ and $\cos(2k\tau_0)$ with details below,
\bal
{\rm oscillation ~ terms} &=
\left[ -\frac{1}{k\tau_0}
+ E_0^2 H_0^2 \tau_0^4
\left( -\frac{13}{40k\tau_0} + \frac{1}{2k^3\tau_0^3} + \frac{3}{8k^5\tau_0^5} \right) \right.
\nonumber \\
& ~~~~~~~~~~~~~
\left.
+ E_0^2 H_0^2 \tau_0^4
\left( \frac{4}{5 k\tau_0} - \frac{7}{6k^3\tau_0^3} - \frac{7}{8k^5\tau_0^5} \right) \cos^2\theta \right]
\sin(2k\tau_0)
\nonumber \\
&+
\left[ -\frac{1}{2k^2\tau_0^2}
+ E_0^2 H_0^2 \tau_0^4
\left( \frac{1}{40} - \frac{9}{16 k^2\tau_0^2} + \frac{3}{16k^6\tau_0^6} \right) \right.
\nonumber \\
& ~~~~~~~~~~~~~
\left.
+ E_0^2 H_0^2 \tau_0^4
\left( - \frac{1}{10} + \frac{4}{3 k^2\tau_0^2} - \frac{7}{16k^6\tau_0^6} \right) \cos^2\theta \right]
\cos(2k\tau_0) ~.
\eal

We can make use of these results in different limits. For our purpose, let us look at the large $k$ behavior and see how the anisotropic part decay as a functions of scale $k$.
Although the oscillation terms also contain interesting properties, such as the scale-independent (to the extend that the non-BD state is valid) anisotropic terms, for reasons we stated previously, we emphasize the anisotropic and non-oscillatory component, in particular the 3rd line in \eqref{Csum_squared}. The naive expectation and the behavior for the BD state, given by the first line in \eqref{Csum_squared} $\sim 1/k^4$, is now changed qualitatively to $\sim 1/k^2$. This is just a special choice of initial state, in principle more variety of scale dependence can arise through this mechanism.

In the usual isotropic inflationary case, the main characteristic effect from the non-BD state is the oscillatory scale-dependence \cite{Brandenberger:2012aj}. This is not because the non-oscillatory part is not affected by the non-BD state; on the contrary, it should be affected, as we can see from the second term in the second line of \eqref{Csum_squared}. The reason is that, in the isotropic case, such a change is observationally indistinguishable from the BD state with similar scale-dependence due to other reasons, like shapes of potential. The effect we considered here is qualitatively different. At the non-oscillatory level which is much easier to observe, this effect is already distinguishable; because it is encoded in a different, anisotropic component, in sharp contrast to the leading, isotropic component.
Furthermore, this is present even in the power spectrum of the single field slow-roll inflation.
In this sense, primordial anisotropy can be a very sensitive probe of the initial state of inflation.

To conclude, relic anisotropies at the onset of inflation induce anisotropies in density perturbations with strong scale dependence. We studied the effects of initial quantum fluctuation states on the details of this scale dependence. We showed that generic quantum states tend to preserve the information of these relics to much shorter scales, than the naive expectation from the background dilution speed of these relics by inflation. Conversely, we can use the detailed scale-dependence of these anisotropies to probe the initial quantum state of our universe.

\section*{Acknowledgments}
XC is supported by the Stephen Hawking Advanced Fellowship. YW is supported by fundings from Kavli IPMU (WPI), the University of Tokyo.

\end{spacing}


\end{document}